\documentclass[10pt,conference]{IEEEtran}
\IEEEoverridecommandlockouts
\usepackage{algorithmic,algorithm}
\usepackage{cite}
\usepackage[utf8x]{inputenc} 
\usepackage{amsmath}
\usepackage{amssymb,amsmath}
\usepackage{graphicx}
\usepackage{epsfig}
\usepackage{grffile}
\usepackage{balance,color}
\providecommand{\keywords}[1]{{\textit{Index Terms}}}
\abovecaptionskip 
\belowcaptionskip 

\begin{document}

\title{SSHealth: Toward Secure, Blockchain-enabled Healthcare Systems}  
\author{Alaa Awad Abdellatif$^{*\dag}$, Abeer Z. Al-Marridi$^{*}$, Amr Mohamed$^{*}$, Aiman Erbad$^{*}$, Carla Fabiana Chiasserini$^{\dag}$,\\ and Ahmed Refaey$^+$ \\ %
$^*$Department of Computer Science and Engineering, Qatar University \\
$^\dag$Department of Electronics and Telecommunications, Politecnico di Torino  \\
$^+$Department of Electrical and Computer Engineering, Manhattan College \\
E-mail:  \{alaa.abdellatif, chiasserini\}@polito.it,  \{aa1107191, amrm, aerbad\}@qu.edu.qa, \\ ahussein01@manhattan.edu \\  %
\thanks{This work was made possible by grant \# QUHI-CENG-19/20-1 from Qatar University. The work of Abeer Z. Al-Marridi is supported by GSRA grant \# GSRA5-1-0326-18026 from the Qatar National Research Fund (a member of Qatar Foundation). The findings  achieved  herein are solely the responsibility of the authors.}
} 
\maketitle

\begin{abstract}

The future of healthcare systems is being shaped by incorporating emerged technological innovations to drive new models for patient care. By acquiring, integrating, analyzing, and exchanging medical data at different system levels, new practices can be introduced, offering a radical improvement to healthcare services.  
This paper presents a novel smart and secure Healthcare system (ssHealth), which,  leveraging advances in edge computing and blockchain technologies,  permits epidemics discovering, remote monitoring, and fast emergency response. 
The proposed system also allows  for secure medical data exchange among local healthcare entities, thus realizing the integration of  multiple national and international entities and enabling the correlation of critical medical events for, e.g., emerging epidemics management and control. In particular, we develop a blockchain-based architecture and enable a flexible configuration thereof, which optimize medical data sharing between different health entities and fulfil the diverse levels of Quality of Service (QoS) that   ssHealth may   require.  
Finally, we highlight the benefits of the proposed ssHealth system and  possible directions for future research.  
\end{abstract}

\section{Introduction\label{sec:Introduction}} 

Developing a smart, efficient and secure healthcare system for improving people's quality of life is a top interest worldwide. A pivotal contribution  to the development of smart-health systems has come from some emerging technologies such as Internet of Things (IoT), Blockchain, and Edge Computing. Advanced e-health applications are expected to inspire fundamental transformations for the healthcare industry towards Healthcare Industry 4.0 (Health 4.0) ‎‎\cite{Health42017}, especially in pre-hospital emergency care situations and for geographically remote areas \cite{NS_Systems}.  
In the age of IoT and Health 4.0, health-related applications are gaining momentum, as the huge amount of data generated through that allows for more in-depth medical studies and patients  feel more secure if their status is precisely monitored even outside the hospital. Moreover, medical data exchange across multiple entities can lead to a better quality level in the care for the patients, improving the response time in  emergency conditions and a more accurate control and management of diseases. 
However, critical challenges have emerged, which need to be faced to ensure high-quality services. Specifically,
\begin{itemize}
	\item	the massive real-time data collected from different health monitoring systems, which need to be efficiently stored, processed, and shared;  
\item	fulfilling diverse security and privacy requirements, while dealing with the complexity of data processing and  transfer;   
\item	ensuring remote accessibility of medical data by different authorized entities to realize  large-scale, low-cost healthcare  and  personalized medicine. 
\end{itemize}
Additionally, it is worth noting that  traditional healthcare systems exhibit  weak security protection and are often subject to attacks:  from 2016 to 2017,  the number of reported healthcare attacks\footnote{https://www.cryptonitenxt.com/resources} increased by 89\%. Thus, using encryption techniques in healthcare systems that provide high security levels is essential, although they may  result in high computational complexity and latency. 

In this work, we argue that building a secure, trusted, and decentralized smart-healthcare system addressing the above challenges can be established leveraging edge computing and blockchain. 
Blockchain is a decentralized ledger of transactions that are shared among multiple entities while preserving the integrity and consistency of the shared data. It is considered as a revolutionary technology that will have a huge impact on the society: in the 2015 World Economic Forum\footnote{https://www.weforum.org/reports/deep-shift-technology-tipping-points-and-societal-impact} report, 58\% of the participants foresaw that 10\% of global Gross Domestic Product (GDP) will be related to the blockchain technology by 2025.  Being decentralized, it well matches the potentiality of edge computing, which can effectively support data storage and processing at different entities as well as their interconnection.  
{ We therefore aim at paving the way to efficient,  blockchain-based, smart-health systems and applications, by answering the following fundamental questions: 
\begin{itemize}
	\item[{\em (i)}] is  blockchain a valid solution  for realizing healthcare systems? 
	\item[{\em (ii)}] and, how can we leverage the blockchain capabilities and the edge computing potentialities to fulfil diverse healthcare applications' requirements?  
\end{itemize}
We address the above aspects as follows:
\begin{enumerate}
	\item we propose a smart, secure, and decentralized healthcare system that relies on blockchain and edge computing technologies to provide a convenient data sharing among multiple entities; 
	\item we formulate a flexible configuration model that enables the blockchain system to support diverse QoS requirements, and we develop an efficient algorithm to solve this model;  	
	\item through numerical results, we show the effectiveness of the proposed approach in improving the blockchain performance for healthcare applications. 
\end{enumerate}
}
 In what follows, we  highlight the advantages of using  blockchain within a healthcare system and present some relevant related work (Section \ref{sec:Sec2}). We then introduce the proposed ssHealth system architecture leveraging edge computing and blockchain, and we present the associated  blockchain configuration model (Section \ref{sec:system}). Finally, we show the benefits of leveraging blockchain and edge computing capabilities within the proposed architecture (Section \ref{sec:Benefits}), and conclude the paper by highlighting  possible directions that are worth to be further investigated (Section \ref{sec:conclusion}). 

\begin{table*}[t!]
	\centering
		\caption{Summary of the relevant work on blockchain in healthcare systems}
	  \label{tab:RelWork}
		\begin{tabular}{|c|c|c|c|} 
			\hline 
		\textbf{Blockchain Type} &\textbf{Description} & \textbf{Limitations} & \textbf{Entities}\\
		\hline
	 	Private (Ethereum) & Blockchain system links patients with doctors   & Latency   &Patients    \\
		\textit{Consensus}: Practical Byzantine & using customized smart contract to record & scalability    & Hospitals   \\
		Fault Tolerance (PBFT)  &  all events on the blockchain &  &\\
		\textit{class}: patient \cite{AB1} &  &   &  \\	
	\hline
		Private (Ethereum) & A blockchain framework is proposed &  Scalability  &Patients    \\
		\textit{Consensus}: Proof of Work (PoW) & for searching encrypted index of Electronic &     & Hospitals  \\
		 \textit{class}: patient \cite{AB2} & Health Records (EHRs) while real data &  & Medical labs\\
		 & stored in database &   &Insurance companies   \\
	\hline
		Private (consortium) &Parallel healthcare system using blockchain, & Latency   &Patients    \\
		\textit{Consensus}: delegated proof of & technology is proposed to link various & scalability    & Hospitals   \\
		stake (DPoS) &  parties for medical data sharing  &security& Healthcare communities \\
		\textit{class}: patient \cite{AB3}   & & & Researchers \\
\hline
			 	Private (Ethereum) &  Blockchain framework is proposed to connect  & Scalability   &Patients    \\
		 \textit{Consensus}: PoW  & the patients with the hospitals to enable &  & Hospitals   \\
		 \textit{class}: patient \cite{AB5} & health-related information exchange   &  & Healthcare institutions \\  
		\hline
	 	Private (Hyperledger fabric) & Blockchain framework is proposed for sharing  & Scalability   &Patients    \\
   \textit{Consensus}:	Byzantine fault-tolerant & processed medical data between  & Patients approval      &	Healthcare providers   \\
		state machine replication & different healthcare entities &  &\\
		\textit{class}: patient \cite{AB11} &  &   &  \\	
		\hline	
		 Private  (Ethereum) &Framework of dual blockchains is proposed, & Storage  &System manager    \\
		\textit{Consensus}: proof of conformance & one to store and share the index of the & scalability    & Hospitals   \\
		\textit{class}: entity \cite{AB12} & EHR with multiple hospitals, and the &  &\\
	 & other to store the original data &   &  \\	
		\hline
		 	Public (Ethereum) & Propose a framework of two coupled blockchains & Latency &Patients    \\
		\textit{Consensus}:	PoW & for managing the storage of two types of  & scalability & Medical institutions   \\
  \textit{class}: entity \cite{AB6} & data to enhance the throughput, accessibility, & computational cost &\\
		 &  and fairness among users &   &  \\
		\hline	
		 	Private (MeDShare) & Blockchain system is proposed to provide  & Privacy   &Patients    \\
		\textit{Consensus}: using consensus nodes & medical data sharing, auditing, and & scalability    & Hospitals   \\
		\textit{class}: patient \cite{AB9} & control over diverse entities &  & Research	institutions \\
		\hline	
Private (Hyper ledger fabric) &Blockchain has been integrated with a tree-based & Privacy   &Patients    \\
		\textit{Consensus}:  voting-based approach &method for medical data sharing between & scalability    & Doctors   \\
		\textit{class}: patient \cite{AB7} &different entities& & Insurance companies \\	
		\hline
		\end{tabular}
\end{table*}

\section{Blockchain for Healthcare Systems \label{sec:Sec2}}

In this section, we discuss the key features of a blockchain-based healthcare systems, also in the light of the recent, related  literature. 
 
\subsection{Why blockchain is needed for healthcare systems? \label{sec:why}}

 A healthcare system comprises diverse organizations, people, and actions whose fundamental role is to monitor, promote, and maintain people's health. It includes, for instance, private clinics, pharmacies, hospitals, health insurance companies, occupational health and safety legislation, as well as the ministry of health. 
Effective e-health systems must provide fast response with high quality service level and security for the entire population, while simultaneously promoting disease prevention and managing costs. To this end, the following issues have to be adequately addressed.  

\textbf{Security and privacy:} 
	 Real-time access to clinical patient's records enables e-health systems to give timely care to the patients through the nearest point of care. Furthermore, healthcare entities may need to share relevant data to provide national first response to epidemics, improved national wide statistics, and enhanced quality of healthcare services. Finally, the dissemination, processing, and analysis of medical data are  crucial for the diagnosis and discovery of new  therapies for  emerging diseases.   
However, medical data exchange across multiple organizations comes with many security and privacy risks; in particular, without effective privacy-preserving schemes in place, users may not accept to share their data with others, which would impair the creation of  a system integrating all healthcare entities. Thus, it is mandatory to provide secure data access and to prevent tracking users' identity as well as raw data disclosure.	 

\textbf{Management of patients' flow:}
While detecting and predicting patients' state through data analytic within one organization maybe possible, managing and correlating patients' related data across multiple entities is quite hard. 
The problem is not due to insufficient resources, but due to insufficient resource management.  
The challenge resides in the ability of healthcare providers to foresee patients' flow, which demands for predictive analytics and collaboration among different entities to align available resources to the forthcoming demand.    

\textbf{Support of diverse QoS requirements:} 
Different E-health applications require diverse QoS requirements, including high data rates, data accessibility anytime and anywhere, low latency, etc. This imposes the need for designing a customized/reconfigurable system that can support diverse applications' requirements.   

Blockchain appears as a perfect solution to all of the above issues. It provides fast, secure exchange and storage of medical data, and it can aggregate different health entities, with diverse policies, and make them part of a unique national healthcare system.    
The power of security in blockchain comes from the collective resources of the crowd, since, most of the entities have to verify each block of data using a consensus algorithm\footnote{Consensus algorithms are mechanisms that ensure the integrity and consistency of the blockchain across all the participating entities  \cite{AB3}.}, e.g. Delegated Proof-of Stake (DPoS) \cite{AB3}. Hence, any cyber attack has to beat the resources of the whole crowd collectively to be able to hack the integrity of the data.  

\subsection{Related work on blockchain-based healthcare systems\label{sec:Related}}
  
{  
Recently, different types of blockchains have been envisioned for the healthcare sector, including public and private  blockchains. Public blockchains offer decentralized and secure data sharing, however, when advanced control and privacy are required, private or permissioned models turn to be more efficient.  
Several blockchain frameworks (e.g., Ethereum and Hyper ledger Fabric), smart contracts\footnote{A smart contract is a software that contains all instructions and rules agreed upon by all the entities to be applied on the blockchain: all the transactions need to be consistent with the smart contract before being added to the blockchain.}, and consensus algorithms have been investigated in the literature.   
The general blockchain architecture mainly consists of: data sender, Blockchain Manager (BM), and verifiers. First, data senders upload their data as ``transactions'' to the nearby BM. Then the BM acts as a verifiers' manager: it  generates unverified blocks, distributes them across the verifiers, triggers the consensus process, and inserts the verified blocks in the blockchain. Hence, the BM acts as the leader, while the verifiers are the followers that cooperate to complete the block verification task. In line with the traditional DPoS consensus scheme, the verifiers take turns to work as BM for a given period of time \cite{BC2019}.      


 For healthcare applications, the blockchain architectures that have been proposed so far can be broadly classified into two categories: patient-based and entity-based. In patient-based architectures, patients participate in the blockchain and transactions are driven by the patient directly. However, such architectures have a limitation in terms of system's scalability. In entity-based architectures, instead, health organizations, hospitals, research institutes, and alike are the main actors, while patients only interact with the health organizations to acquire the service they need.      
 According to our survey, 83\% of the systems proposed  since 2016 are patient based, while 17\% are entity based. Table~\ref{tab:RelWork} reports recent works in this area, highlighting the  encryption techniques and consensus algorithms they adopt, as well as some of the limitations they exhibit.  }    
{  In particular, several approaches suffer from poor scalability and slow response. Being swift response a major goal for emergency care, some studies aim to overcome these limitations using: (i) an external database at the data owner \cite{AB2}, which stores the raw data and shares the index to the data in the blockchain, (ii) a private blockchain and a consortium blockchain \cite{AB12}, where the private blockchain is responsible for storing the data while the consortium blockchain stores the index of the data. However, allowing  data storage  at a single entity comes with the risk of a single point of failure, while considering two types of blockchain may have an impact in terms of  computational cost. }  
We therefore envision a solution that combines a blockchain-enabled architecture with intelligent processing at the edge, so as to support fast, secure, and scalable exchange of medical data.

\begin{figure*}[htp]
	\centering
		\scalebox{3.6}{\includegraphics[width=0.27 \textwidth]{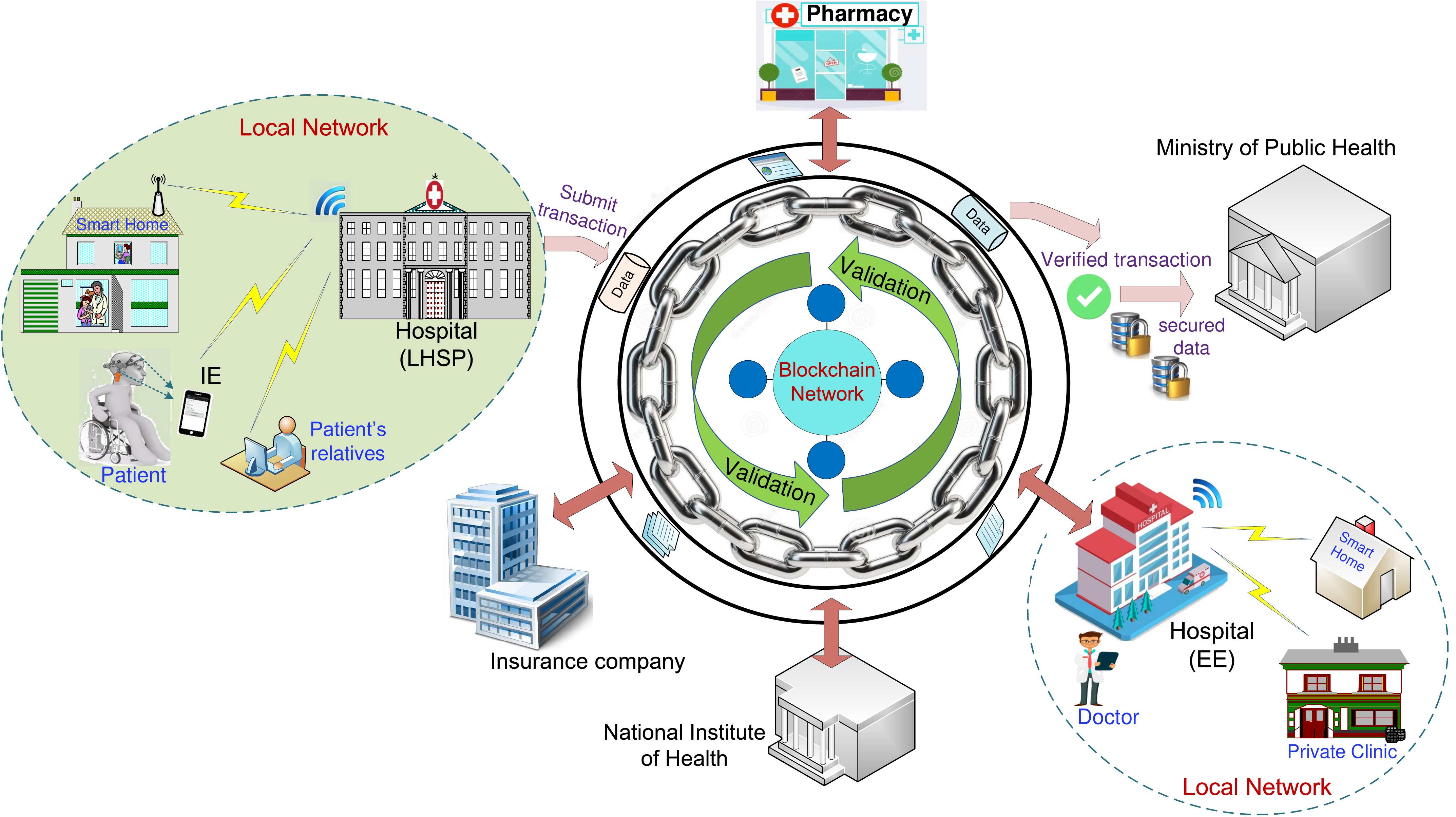}}
	\caption{Proposed ssHealth system architecture. }
	\label{fig:system_model}
\end{figure*}

\section{ssHealth Architecture and Blockchain Configuration\label{sec:system}}

We now describe the proposed ssHealth system architecture, then we discuss the blockchain approach we adopt. Finally, we present a method for optimally configuring the blockchain system so that it effectively addresses the challenges and requirements posed by diverse e-health applications.

\subsection{ssHealth architecture\label{sec:Architecture}} 

The proposed system architecture, shown in Figure~\ref{fig:system_model}, includes two main network sub-systems: (a) a Local network and (b) a Blockchain network. 
{  For the sake of scalability, it is assumed that healthcare entities collect health-related information from the local network, process these data, and share important information through the blockchain network. The shared data are validated and  stored locally by the different entities in the blockchain, which are trusted entities with large storage and computational capabilities \cite{suggested}. }

The local network stretches from the data sources located on or around  patients to the Local Healthcare Service Provider (LHSP), e.g., a hospital. It contains the following major components:  \\ 
\textit{a.1) Internet of Medical Things (IoMT):} A combination of sensor nodes attached/near to the patients to be leveraged for monitoring health conditions and activities within the smart assisted environment. Examples include: body area sensor networks (i.e., implantable or wearable sensors that measure different biosignals and vital signs),  smartphones, IP cameras, and external medical and non-medical devices. \\       
\textit{a.2) Internal Edge (IE):}  This edge node implements local processing functions between the data sources and the LHSP.  
Specifically, the IE analyzes the gathered medical and non-medical data from different sources, obtains the information of interest, and forwards the processed data/extracted information to the LHSP. Moreover, IE can be a mobile node (e.g., a smartphone) or an infrastructure edge node (e.g., a wireless router or an access point).    
Importantly, the IE can  optimize the medical data delivery based on the context (i.e., data type, supported application, and patient's state) as well as on the conditions of wireless connectivity.     
Furthermore, different specialized healthcare applications can be implemented at the IE to allow  patients to actively participate in their treatment and ubiquitously interact with their doctors anytime and anywhere.  \\ 
\textit{a.3) Local Healthcare Service Provider (LHSP):} 
An LHSP can be a hospital, which monitors and provides the required healthcare services for the local patients, records the patients' state,  and puts in place fast emergency services if needed. Importantly, the LHSP plays a significant role in monitoring of patients' state not only inside the hospital (intra-hospital patient care), but also outside  (e.g., home patient care). Also, it can be connected with the private-local clinics that may transfer patients to it for more advanced care, or even with patient's relatives to follow up on the patient's conditions. 

 As far as the blockchain network is concerned  (see Figure~\ref{fig:system_model}), the core is the blockchain-based data sharing architecture that enables secure access, processing, and sharing of medical data among  healthcare entities.   
Blockchain is suitable for secure medical data sharing because of its immutability and decentralization features, which are perfectly consistent with our proposed ssHealth architecture. Using blockchain, all transaction blocks (i.e., containing health-related information) can be securely shared, accessed, and stored by physicians, decision makers, and other healthcare entities.   
The latter ones include, but are not limited to:   \\   
\textit{b.1) External Edge (EE):}   
In the proposed architecture, a hospital or an LHSP have more advanced tasks than the ones mentioned above: it can act as an EE that is responsible for data storage, applying sophisticated data analysis techniques, population health management, and sharing important health-related information with public health entities. Hence, leveraging the power of edge computing, each entity can verify the authenticity and integrity of the medical data at the EE before sharing it within the blockchain. \\ 
\textit{b.2) Insurance companies:} 
 One important aspect for e-health systems is integrating healthcare providers, patients, and payers into one ``digitized community", in order to improve quality of services and drive costs  down. Indeed, to realize a sustainable healthcare-business model, healthcare providers will have to own health plans powered by  insurance companies. \\  
\textit{b.3) Pharmacies:}
The main pharmacies' duties include processing prescriptions, storing and providing access to disbursed prescriptions, and ensuring patients' privacy. On top of it, pharmacies have to coordinate with private insurance companies to submit insurance claims, ensure payment, and resolve denials of coverage. Pharmacies may also communicate with prescribers to confirm the dosage and formulation (e.g., liquid or tablet), or to replace prescribed brand name with a generic equivalent. Thus, it is crucial to have a secure communication system to exchange such information with different associated entities.   \\  
\textit{b.4) National Institutes of Health (NIH):}
NIH are major players in clinical research and health education. The latter in particular  is a process in which all public healthcare institutes, hospitals, and medical care personnel are involved. Thus, NIH should cooperate with healthcare service providers to develop joint educational programs and services for pursuit scientific research and preventive medicine.  \\   
\textit{b.5) Ministry of Public Health (MOPH):}
The main role of MOPH is monitoring the quality and effectiveness of healthcare services through coordination with different health entities. MOPH waives the responsibility of healthcare services to the hands of public and private health sectors while regulating, monitoring, and evaluating their healthcare services to guarantee an acceptable quality of care level. Thus, MOPH is committed to establishing an environment that promotes high-quality services by sharing relevant information with its partners such as health insurance companies.  

\begin{figure*}
	\centering
		\scalebox{3.25}{\includegraphics[width=0.27 \textwidth]{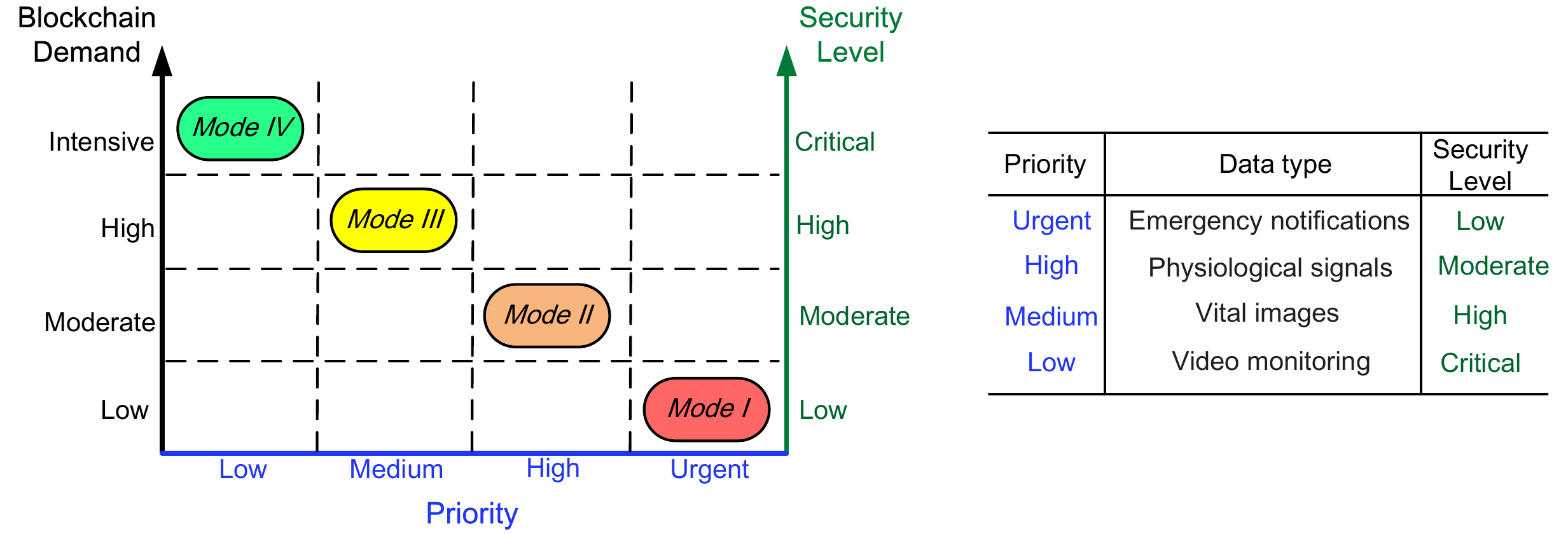}}
	\caption{Blockchain modes based on the data priority and required security level.}
	\label{fig:Edgetasks}
\end{figure*}

\subsection{Optimal blockchain configuration \label{sec:Cases}}  

Leveraging the above ssHealth architecture, we develop a blockchain-based data sharing scheme that enables medical data access, processing, and sharing among the aforementioned healthcare entities. However, blockchain poses  a new challenge, i.e., finding the optimal  trade-off among security level, latency, and cost. 
 Indeed, due to the need of coordinating the transactions of multiple entities, public blockchain is slower than traditional databases, implying a service latency that may be unacceptable for several applications (e.g., emergency management).  
We address this challenge by designing  a priority-based secure data sharing scheme, as detailed below.  

We draw on the BM concept (see Section \ref{sec:Related}), i.e., a logical role that any entity in the proposed architecture can take on, possibly by taking  turns, or that can be taken by the EE that wants to share its data.  
In particular, in our scheme the BM  is  in charge of: 
(i) collecting the transactions received from the different entities, (ii)  preparing and distributing unverified blocks to the verifies  (e.g., hospitals, NIH, and MOPH, which have sufficient computation and storage resources), (iii) updating blockchain configuration considering urgency and security level of the collected data, and (iv) interacting with the verifiers to complete the block verification tasks.  
  BM is thus a critical component, which should  carefully select the blockchain configuration in terms of number of verifiers and number of transactions per block.  These parameters should be dynamically set based on the diverse applications' requirements and data types, and in such a way  that  the optimal trade-off among security, latency, and cost is established. 
As an example, Figure~\ref{fig:Edgetasks} illustrates  the case where high-priority data are received requiring minimum security, e.g., emergency notifications, and should be  dealt with a restricted blockchain, i.e., minimum number of verifiers. On the contrary, for low priority types of data but requiring a high security level (such as video monitoring), fully restricted blockchain mode should be used. In general, the more verifiers participate in the block verification stage, the higher the security level is, but also the larger the latency (due to the verification delay) and the higher the cost (due to verification fees) that are experienced \cite{BC2019}, \cite{moreverifier}. Instead, as the number of transactions per block grows, the latency increases, while the  cost per transaction decreases.  
We also remark that data types and priorities are defined at the edge by applying different data classification, event detection, and summarization techniques.

As a case study, we focus on private blockchain framework with DPoS consensus scheme, which performs the consensus process using pre-selected verifiers with moderate cost. Also, we consider that the BM resides at the EE and has to: 
	(i) detect the patient's context (including patient's conditions, data type, and security requirements), and 	
	(ii) map the patient's context into different configuration modes of the blockchain.
To represent the different conflicting metrics the BM has to play with, namely,  latency ($L$), security ($S$), and cost ($C$), we define an aggregate utility $U$, which combines them into a single function:
 \begin{equation}
U = \alpha \cdot \frac{L}{l_m} + \beta \cdot \frac{s_m}{S} + \gamma \cdot \frac{C}{c_m},  
\label{eq:utility}
 \end{equation} 
where $\alpha$, $\beta$, and $\gamma$ are weighting coefficients that represent the relative importance of the considered metrics, such that $\alpha+\beta+\gamma=1$.  
However, these metrics has different values and units, which must be normalized with respect to their maximum value (denoted by $l_m$, $s_m$, and $c_m$, respectively) to make them comparable.

The BM can then set the best blockchain configuration, by solving the following problem:
%
\begin{equation}
\begin{aligned}
& \underset{m, \theta}{\text{minimize}}
& & \alpha \cdot \frac{L}{l_m} + \beta \cdot \frac{s_m}{S} + \gamma \cdot \frac{C}{c_m} \\
& \text{subject to}
& & v \leq m \leq M, \\
&&& t \leq \theta \leq N,
\label{optimizefinal}
\end{aligned}
\end{equation}
where $m$ is the number of selected verifiers, with  maximum and minimum values equal to $M$ and $v$, respectively, and $\theta$ is the number of transactions per block, with  maximum and minimum values equal to $N$ and $t$, respectively. 
In (\ref{eq:utility}), the security level is defined as $ S = {\kappa \cdot m^q}$, where $\kappa$ is a coefficient given by the system, and $q\geq 2$ is an indicator factor representing the network scale \cite{BC2019}.  
{  
$L$ refers to the verification latency, which includes the four steps of the block verification process: (i) unverified block transmission from the BM to the verifiers, (ii) block verification time, (iii) verification result broadcasting and comparison among verifiers, and (iv) verification feedback transmission from the verifiers to the BM. Hence, the normalized latency is defined, according to \cite{BC2019}, as  
\begin{equation}
L = { \frac{\theta \cdot B}{r_d}+ \max_{i \in \left\{v, \cdots, M\right\}} \left({\frac{K}{x_i}}\right) + \psi \cdot \theta \cdot B \cdot m + \frac{O}{r_u} },
\label{eq:latency}
\end{equation}
where $B$ is the transaction size, $K$ is the required computational resources for block verification task, $x_i$ is the amount of available computational resources at verifier $i$, ${O}$ is the verification feedback size, and $r_d$ and $r_u$ are, respectively, the downlink and the uplink transmission rate, from the BM to the verifiers and vice versa. In (\ref{eq:latency}), $\psi$ is a predefined parameter that can be defined leveraging the statistics on previous processes of block verification (as detailed in \cite{BC2019}).   
} 
%
Finally, the cost function is defined as 
$  C = \frac{\sum_{i=1}^{m} c_i}{\theta}$, 
where $c_i$ is the computational cost of verifier $i$, given by $c_i = \rho_i \cdot x_i$. Therein, $\rho_i$ represents the payment from verifier $i$ to a cloud service provider, in order to acquire the needed resources for the verification process.  
     
By defining the weighting coefficients $\alpha$, $\beta$, and $\gamma$ as functions of data types and application's requirements, the optimal number of verifiers $m^*$ and transactions per block $\theta^*$ are obtained by solving (\ref{optimizefinal}). However, the above optimization problem is an integer programming optimization, which is an NP-complete problem \cite{a15}.    
In light of the problem complexity, we propose Algorithm \ref{alg:alg_BMS} for an efficient and swift solution. In this algorithm, verifiers are selected in an ascending order based on their associated latency, i.e., those verifiers that finish block verification faster will be selected first.  

\begin{algorithm}
\caption{Blockchain Mode Optimization}
\label{alg:alg_BMS}
\begin{algorithmic}[1]
\STATE {\textbf{Input:} $x_i$, $\rho_i$, $v$, $M$, $t$, $N$.}
\STATE {Initially: set $m=v$, $\theta=t$, and compute $U$ as in (\ref{eq:utility}).}
\FOR {$m = v:M$}
\FOR {$\theta = t+1:N$}
\STATE {Update $U$ based on (\ref{eq:utility}).}
\IF {$U(\theta) > U(\theta-1)$}
\STATE {$\theta^*=\theta-1$.}
\STATE Break \% $\theta^*$ is obtained 
\ENDIF
\ENDFOR
\IF {{$m>v$} \& {$U(m) > U(m-1)$}}
\STATE {$m^*=m-1$.}
\STATE Break \% $m^*$ is obtained
\ENDIF
\ENDFOR
\STATE {\textbf{Output:} $m^*$, $\theta^*$.}
\end{algorithmic}
\end{algorithm}

\begin{figure}[t!]
	\centering
		\scalebox{1.6}{\includegraphics[width=0.27 \textwidth]{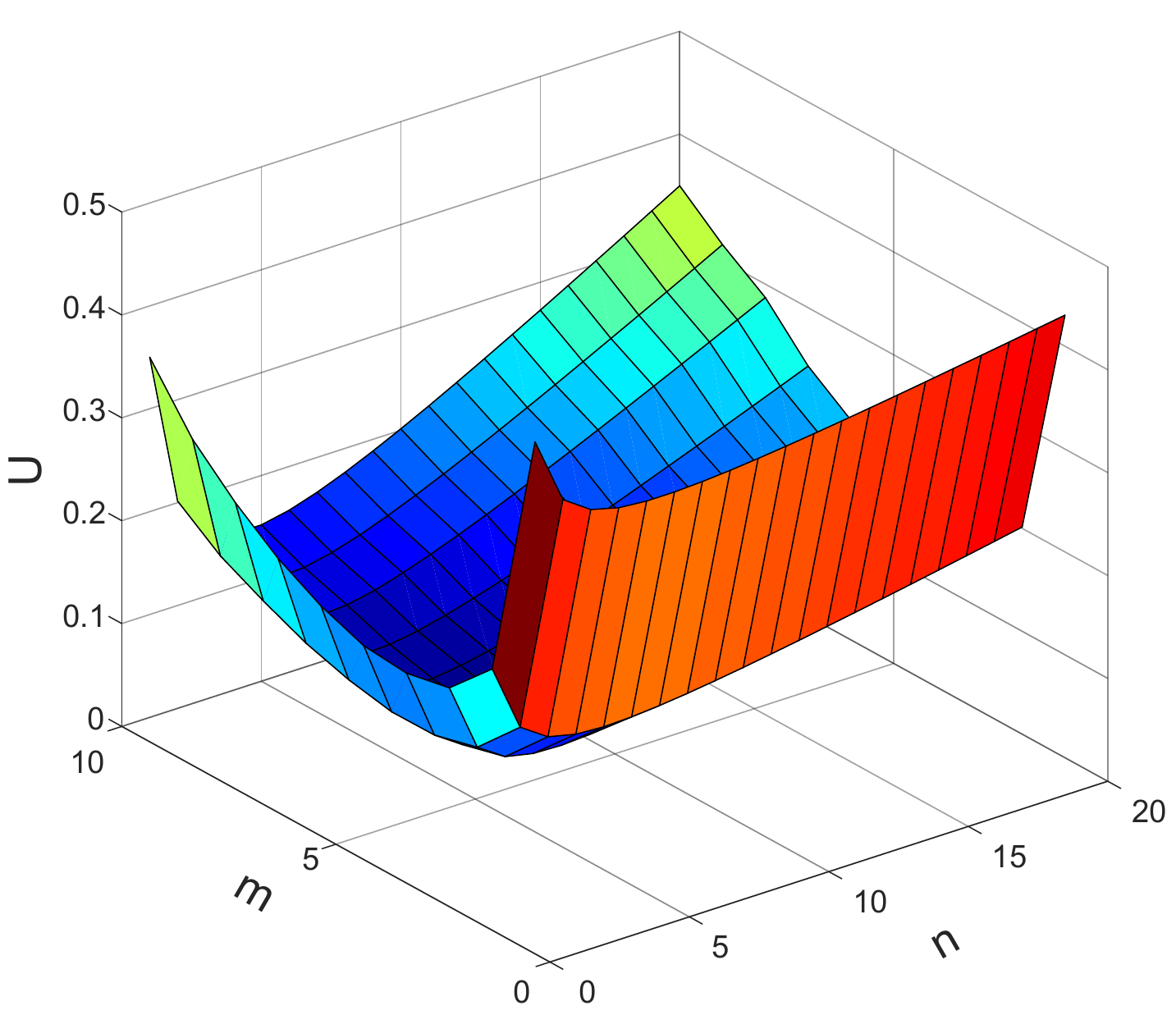}}
	\caption{The proposed utility function as the number of verifiers ($m$) and the number of transactions per block ($\theta$) vary. }
	\label{fig:BlockchainTradeoff}
\end{figure}

\begin{figure}[t!]
	\centering
		\scalebox{1.7}{\includegraphics[width=0.27 \textwidth]{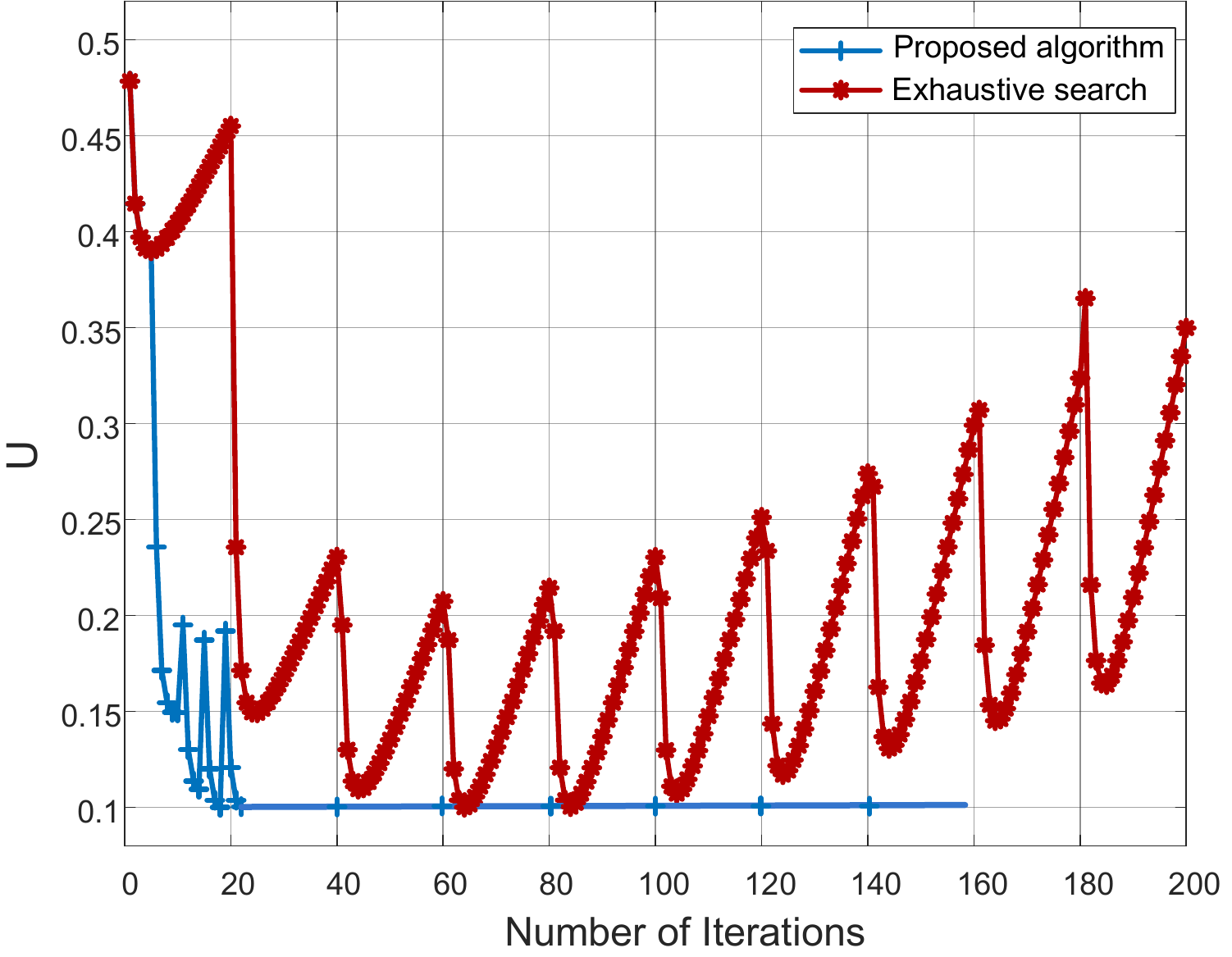}}
	\caption{Convergence behavior of the proposed algorithm compared to the solution obtained through exhaustive search. }
	\label{fig:BlockchainModes}
\end{figure}

{ 

\begin{table}[t!]
	\centering
		\caption{Simulation Parameters}
	  \label{tab:SysParameters}
		\begin{tabular}{|c|c||c|c|} 
			\hline 
		\textbf{Parameter} & \textbf{Value} & \textbf{Parameter} & \textbf{Value}\\
		\hline
		$M$ &	 10 & $N$ &	 20 \\
		\hline
	 	 $v$ &  2 & $t$ &  2 \\
		\hline
		$r_d$ &   1.2 Mb/s & $r_u$ &	 1.3	Mb/s   \\	
		\hline
		$O$ &	 0.5 Mb	& 	$B$ &	 1 kb	 \\
		\hline
		\end{tabular}
\end{table}
}
Figure~\ref{fig:BlockchainTradeoff} depicts the variations in the objective $U$ as the number of verifiers $m$ and the number of transactions per block $\theta$ vary, for  applications with similar requirements in terms of security, latency, and  cost ($\alpha=\beta=\gamma$). Other simulation parameters are reported in Table \ref{tab:SysParameters}.  
 Furthermore, Figure~\ref{fig:BlockchainModes} shows the convergence behavior  of the proposed algorithm to the optimal solution obtained by exhaustive search (or brute-force search) \cite{exhaustivesearch}, for  $M=10$ and $N=20$. We observe that our algorithm requires only 23 iterations to reach the optimal solution compared to exhaustive search that still does not converge after 200 iterations.    

\section{Benefits of ssHealth \label{sec:Benefits}}

Given the requirements of e-health applications discussed in Section \ref{sec:Sec2}, we  now highlight the effectiveness of our ssHealth system in fulfilling such requirements.     \\ 
\textbf{Privacy and security:} 
Slicing the overall system into local and blockchain network facilitates medical data processing, accessing, sharing, and storage while dealing with encrypted data throughout the entire process.   
The proposed ssHealth system enhances accessibility and information sharing between patients and hospitals to provide effective and safe healthcare services, while protecting healthcare systems from cybersecurity threats. It also allows for preventing privacy threats that data sharing entails leveraging blockchain technology, which provides a finite number of authorized entities with secure access to patients' health records by embedding access control rules in the smart contracts. Moreover, edge computing capabilities enable each entity (at the EE) to verify the authenticity and integrity of the medical data before sharing it within the blockchain network. 

{  We remark that designing novel privacy-preserving schemes for the blockchain is not within the scope of this paper, but the proposed architecture can accommodate different  techniques at both the IE and the EE, in order to ensure the required privacy level.}   \\       
\textbf{Scalability and management:} 
 For implementing an effective healthcare system, various entities should collaborate, and a global health system should be created. The proposed ssHealth system realizes such collaboration efficiently by: splitting the workload among different entities, enabling secure data exchange, avoiding the hurdles of managing the available resources or data warehouse.  
Furthermore, it enables building a scalable and reliable healthcare system by: (i) connecting different physician groups and health entities, which facilitates the implementation of clinically-integrated, high-value networks for better patient care; (ii) enabling secure medical data sharing, which helps health institutes and pharmacies to anticipate and manage resources across their health systems (e.g., hospital capacity and drugs).    \\ 
\textbf{Fulfilling diverse QoS requirements:} 
The proposed ssHealth system can not only transfer massive amounts of data securely, but also analyze data efficiently at the EE to extract meaningful and concise information to be shared with the other entities.  Moreover, it efficiently supports different types of applications and data according to their QoS requirements, e.g., demands for high data rates and swift response. 

At last, we remark that the proposed system allows for improved healthcare services by developing a patient-centric, physician-aligned healthcare management model. Such architecture can be leveraged to avoid visits to the hospital emergency wards in non-critical situations, thus reducing costs and improving health-case services for patients with serious conditions. 


\section{Conclusions and Future Directions\label{sec:conclusion}}

In this paper, we envisioned a novel e-health system for creating effective, large-scale and collaborative systems able to  provide high-quality patients' care and to make significant advancements in disease  treatments  through secure data sharing. The proposed ssHealth system integrates edge computing and blockchain to enable the exchange of large amount of medical data generated by different healthcare entities, while preserving the patients' privacy.     
 Additionally, we defined a novel mechanism that can be implemented within the blockchain network to ensure fast response, scalability, and secure transmission of medical data. It is shown that  mapping the characteristics of the collected data onto appropriate configurations of the blockchain can significantly enhance the performance of the overall ssHealth system, while satisfying diverse applications' requirements.  

In this context, several promising directions for  future research emerge, which include: \\ 
{\em (i) Developing various cybersecurity  schemes at the IE and EE to achieve a robust privacy protection of medical data and patients' profiles.}  Maximizing security level for health applications may substantially degrade QoS and cause service disruption. Thus, considering the concept of quality of protection (QoP) while providing security and privacy  is mandatory. In this regard, developing QoP-aware schemes can ensure  different levels of  anonymity and privacy, and optimize misbehavior detection and encryption, according to the type  of the collected data and the level of emergency of the  situations we have to deal with. \\ 
{\em (ii) Further optimizing the blockchain parameters, such as block size, transaction size, and number of blockchain channels.} With the evolution of the blockchain, new features have been added for enhancing security and scalability. One important feature is the multi-channel blockchain network, where each channel corresponds to a separate chain of transactions and can be used for enabling data access and 
private communications among the channel users. By leveraging such a concept, specific geographical areas or group of patients and hospitals can share their data, thus ensuring secure data access and system scalability.   

\balance 

\bibliographystyle{IEEEtrannames}
\bibliography{ssHealth}

\end{document}